\documentclass[
	reprint,
	groupedaddress,
	amsmath,amssymb,
	prb,
	floatfix,
	superscriptaddress,
	twocolumn,
	aps,
]{revtex4-2}

\usepackage[utf8]{inputenc} %
\usepackage[T1]{fontenc}    %
\usepackage{helvet}         %
\usepackage{mathpazo}       %
\usepackage{microtype}      %
\usepackage{amsmath}
\usepackage{bm}
\usepackage{amsfonts}
\usepackage{amssymb}
\usepackage{amsthm}
\usepackage[caption=false]{subfig}
\usepackage{graphicx}
\usepackage{hyperref}
\usepackage{cleveref}
\usepackage[T1]{fontenc}
\usepackage{multirow}
\usepackage{lipsum} %
\usepackage{color,soul}
\usepackage[export]{adjustbox}
\usepackage{natbib}

\usepackage{todonotes}

\begin{document}

\title{Large-scale Monte Carlo/Molecular Dynamics simulations with Atomic Cluster Expansion potentials in LAMMPS}

\author{Lorenzo Piersante}
\email{lorenzo.piersante@epfl.ch}
\affiliation{Laboratory of Materials Design and Simulation (MADES), Institute of Materials, \'{E}cole Polytechnique F\'{e}d\'{e}rale de Lausanne}

\author{Anirudh Raju Natarajan}
\email{anirudh.natarajan@epfl.ch}
\affiliation{Laboratory of Materials Design and Simulation (MADES), Institute of Materials, \'{E}cole Polytechnique F\'{e}d\'{e}rale de Lausanne}

\date{\today}

\begin{abstract}
     Monte Carlo/molecular dynamics (MCMD) simulations are widely used to study the properties of materials at the mesoscale. Simulations with hundreds of thousands of atoms usually rely on empirical potentials because MCMD with machine-learning interatomic potentials is impractical at this scale in existing implementations. Here, we introduce a hybrid MPI/OpenMP scheme for MCMD with Atomic Cluster Expansion (ACE) potentials in LAMMPS. Monte Carlo moves are evaluated from local energy changes and carried out in parallel over MPI ranks, while molecular dynamics is parallelized over OpenMP threads. At a fixed number of CPUs, the wall time of the Monte Carlo moves decreases by a factor equal to the number of MPI ranks. Larger simulation cells also reduce the statistical errors of free energy calculations because thermal fluctuations decrease with system size. We demonstrate the scheme by calculating free energies of the HCP and BCC phases in the strongly anharmonic Hf--Ti system and by simulating zirconium segregation in a nanocrystalline copper sample of 670,000 atoms. For Cu--Zr, an ACE potential and an empirical potential predict similar grain boundary compositions but different amounts of zirconium dissolved in the grains. The scheme makes MCMD simulations with machine-learning interatomic potentials practical at the length scales of microstructure.
\end{abstract}

\maketitle

\section{Introduction} \label{sect:introduction}
Molecular dynamics (MD) is widely used to study atomistic processes in materials. Its combination with Monte Carlo (MC) sampling in Monte Carlo/molecular dynamics (MCMD) simulations underpins both free energy calculations and property prediction at the microstructural length scale. These simulations have been applied to precipitation \cite{starikov_efficient_2026}, grain boundary complexions \cite{pan_effect_2016,sansoz_disordered_2026}, and phase diagram calculations \cite{nitol_moment_2025,miryashkin_clarifying_2026}. Empirical potentials such as the embedded atom method (EAM) and angular-dependent potentials are commonly used in MCMD owing to their efficiency \cite{starikov_angular-dependent_2026}, but their physical fidelity is often insufficient to capture subtle effects \cite{moller__2018}.

Starting with Behler--Parrinello neural networks \cite{behler_generalized_2007}, machine-learning interatomic potentials (MLIPs) have given access to time and length scales inaccessible to \textit{ab initio} MD while retaining near \textit{ab initio} accuracy. However, the computational cost of MLIPs remains considerably higher than that of empirical potentials \cite{leimeroth_machine-learning_2025}. While this cost is acceptable for standard MD, it is a challenge for MCMD because the MC routines in commonly used simulation codes such as LAMMPS \cite{thompson_lammps_2022} are optimized only for EAM potentials. By default, these routines calculate the energy of the entire simulation box before and after each MC move, although the finite interaction radius of most interatomic potentials would allow local energy evaluations.

Recently, Menon \textit{et al.} \cite{menon_computing_2026} introduced a modified version of the MC routines in LAMMPS for Atomic Cluster Expansion (ACE) potentials \cite{lysogorskiy_performant_2021,bochkarev_efficient_2022}. Their implementation exploits local energy evaluations to reduce the cost of each MC move, but it runs only serially and is therefore impractical for large systems. Sadigh \textit{et al.} \cite{sadigh_scalable_2012} addressed the challenge of large-scale MCMD by proposing a parallel MC algorithm for sampling the semigrand canonical (SGC) and variance-constrained semigrand canonical (VC-SGC) ensembles. However, the LAMMPS implementation of this algorithm accepts only EAM potentials, so large-scale MCMD with MLIPs remains impractical.

In this work, we address this limitation by combining the parallel MC algorithm of Sadigh \textit{et al.} \cite{sadigh_scalable_2012} with the local evaluation of ACE potentials. We use hybrid MPI/OpenMP parallelization to accelerate MC without slowing down MD. Together, these two elements allow MCMD simulations of systems containing hundreds of thousands of atoms at near \textit{ab initio} accuracy. Equilibrium microstructures predicted by empirical potentials and MLIPs can then be compared directly. Free energy calculations also benefit from this scheme because the smaller thermal fluctuations of larger systems reduce the statistical errors. We demonstrate the scheme with ACE potentials by calculating free energies of the HCP and BCC phases in the strongly anharmonic Hf--Ti system and by examining zirconium segregation in nanocrystalline copper.

\section{Sampling chemical configurations with Monte Carlo/molecular dynamics} \label{sect:sampling_chem}
Domain decomposition and parallel execution allow MD to scale to large systems. Large-scale MC simulations require the same approach, but the parallelization strategy depends on the thermodynamic ensemble. Systems at fixed composition are normally simulated in the canonical ensemble, where identity-swap moves sample the chemical configurations. Each move swaps the species of two randomly selected atoms and is accepted according to the Metropolis criterion \cite{metropolis_fast_1953}. The canonical ensemble is difficult to parallelize because each swap requires atomic species to be communicated across computational subdomains. This restricts the size of the systems that can be simulated efficiently \cite{sadigh_scalable_2012}. The limitation is lifted in the SGC ensemble, where the exchange chemical potential is fixed and the composition equilibrates through identity-exchange moves. In a binary $A$--$B$ system, an identity-exchange move changes the species of a randomly selected atom and is accepted with probability
\begin{equation}
    \mathcal{A}_{SGC} = \text{min} \left \lbrace 1, \exp[-\beta(\Delta U \pm \Delta \mu)] \right \rbrace
    \label{eq:sgc_ar}
\end{equation}
where $\Delta U$ is the potential energy change of the MC move, $\Delta \mu = \mu_A - \mu_B$ is the exchange chemical potential, and $\beta = 1/k_B T$. The sign before $\Delta \mu$ is positive when the new species is $B$ and negative otherwise. An identity-exchange move is a local operation because, for interatomic potentials with a finite interaction radius, $\Delta U$ depends only on the local environment of the selected atom. SGC MC moves can therefore be carried out independently in different regions of the simulation box, provided that they do not affect each other \cite{sadigh_scalable_2012}.

However, the SGC ensemble cannot access compositions that lie within two-phase regions of the phase diagram \cite{sadigh_calculation_2012}. The VC-SGC ensemble removes this restriction by introducing a quadratic bias potential $\Delta U_{VC}$ that sets a target mean composition and the magnitude of the composition fluctuations. The effective potential energy is then $U_{VC} = U + \Delta U_{VC}$ \cite{sadigh_scalable_2012, sadigh_calculation_2012}, with
\begin{equation}
    \Delta U_{VC} = \kappa \left (N c + \frac{\phi}{ 2 \kappa} \right )^2
    \label{eq:bias_potential}
\end{equation}
where $N$ is the number of atoms in the system and $c$ is the atomic fraction of species $B$. The parameter $\phi$ determines the mean composition, while $\kappa$ controls the magnitude of the composition fluctuations.

The bias potential $\Delta U_{VC}$ makes the acceptance of an identity-exchange move dependent on the global composition. To maintain a parallel implementation, the parameter $\phi$ is written as
\begin{equation}
    \phi = \Delta \mu_{\text{guess}} - 2 \kappa N c_0
    \label{eq:phi_vc}
\end{equation}
where $c_0$ is the target composition and $\Delta \mu_{\text{guess}}$ is a guess exchange chemical potential. With this choice, the change in $U_{VC}$ caused by a set of identity-exchange moves with total composition change $\Delta c$ separates into a local term and a remainder that depends on the global composition. Therefore, the MC routine involves two stages \cite{sadigh_scalable_2012}. The first stage treats the local term with one independent SGC MC move in each computational subdomain, using \cref{eq:sgc_ar} with $\Delta \mu = \Delta \mu_{\text{guess}}$. The second stage accounts for the remainder by collectively accepting or rejecting these moves with probability
\begin{equation}
    \mathcal{A}_{VC}^{\text{glob.}} = \text{min} \left \lbrace 1, \exp \lbrace -\beta \kappa N^2 [\Delta c^2 + 2 \Delta c (c_i - c_0)]\rbrace \right \rbrace
\end{equation}
where $c_i$ is the composition before the MC routine.

\section{Hybrid MPI/OpenMP parallelization} \label{sect:parallelization}
\begin{figure*}[!ht]
    \centering
    \subfloat[][\label{fig:domain_boundary}]{\includegraphics[width=0.33\linewidth]{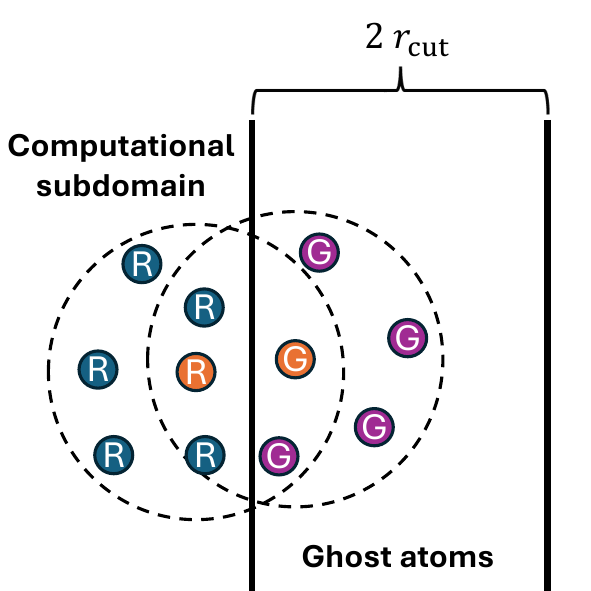}}
    \subfloat[][\label{fig:mpi_omp_parallelization}]{\includegraphics[width=0.33\linewidth]{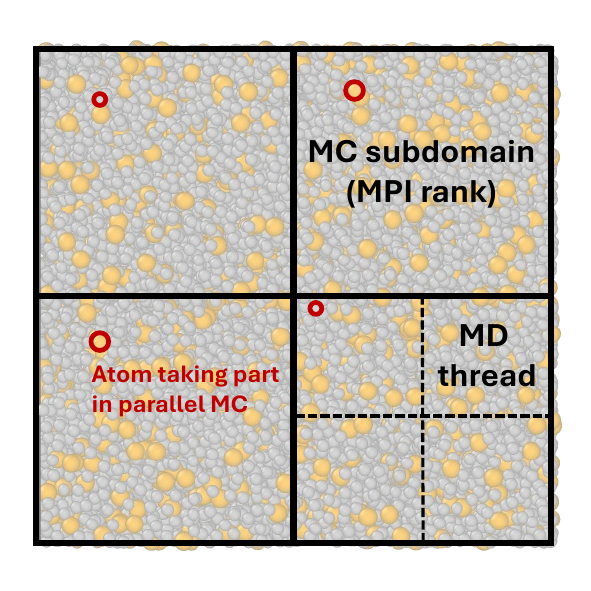}}
    \subfloat[][\label{fig:scaling}]{\includegraphics[width=0.33\linewidth]{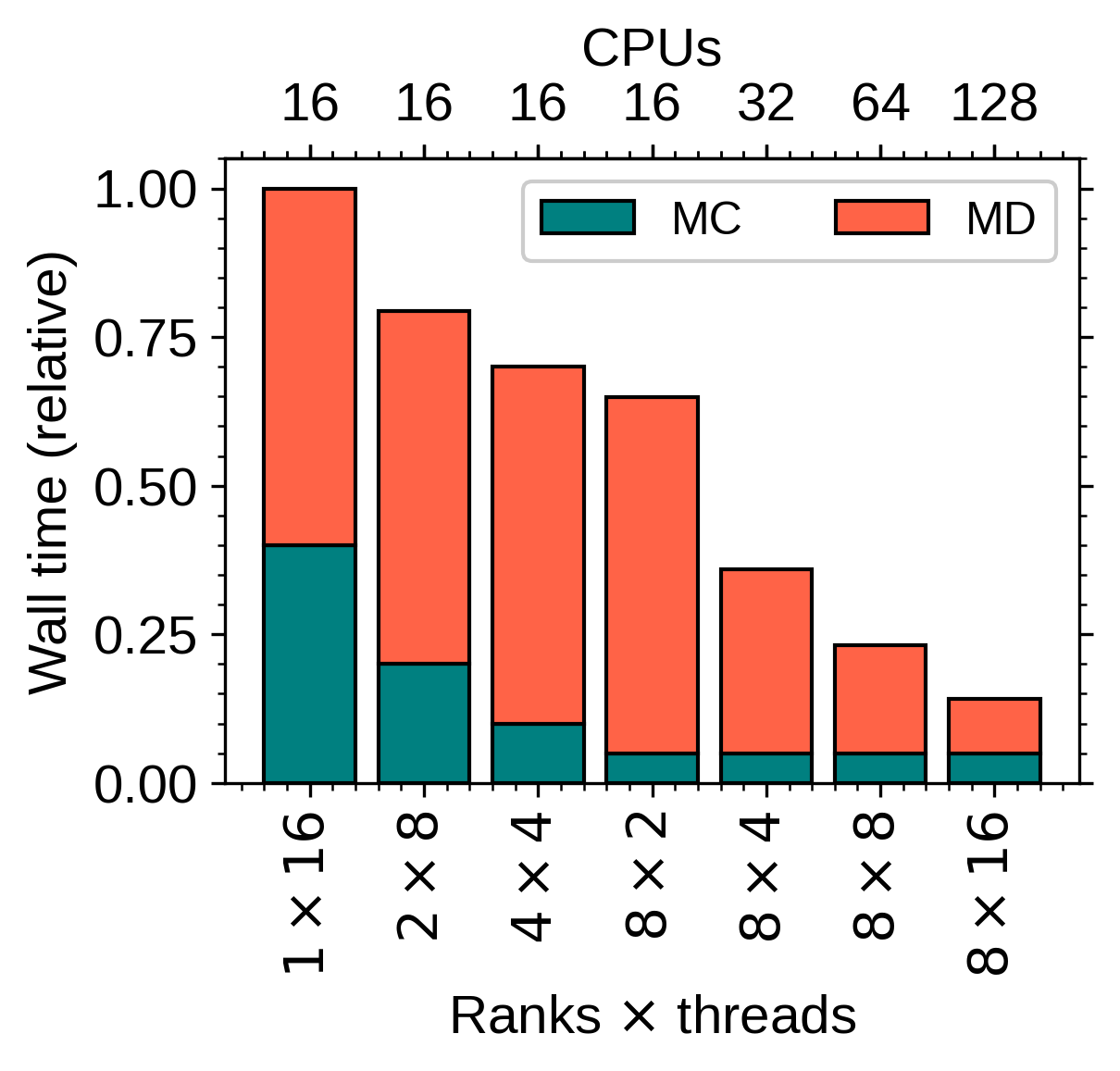}}
    \caption{(a) Neighborhood of an atom close to a LAMMPS subdomain boundary. Dashed circles mark the neighborhoods of the two orange atoms. The neighborhood of the orange real atom (R) contains real atoms and ghost atoms (G). The neighborhood of the orange ghost atom contains further ghost atoms, so the ghost-atom region must span at least twice the interaction radius $r_{\text{cut}}$ for local energies to be evaluated correctly. (b) Schematic of the hybrid parallelization in LAMMPS. Each subdomain corresponds to one MPI rank and carries out its own MC moves on atoms such as those circled in red. The MD within each subdomain is parallelized at a finer level over OpenMP threads (dashed lines). (c) Relative wall time of VC-SGC MCMD simulations of liquid Mg--Nd at 1100\,K and $c_{\text{Nd}} = 0.1$ for different combinations of MPI ranks and OpenMP threads. The bottom axis gives the number of ranks and the number of threads per rank, while the top axis gives the total number of CPUs. Wall times are relative to the configuration with one rank and 16 threads. The MC wall time depends only on the number of MPI ranks, while the MD wall time depends on the total number of CPUs.}
\end{figure*}
MCMD simulations with MLIPs require efficient use of computational resources. By default, LAMMPS evaluates the energy of the entire simulation box before and after each MC move, so the cost of a move grows with system size. To address this issue, we devised a hybrid MPI/OpenMP scheme that brings million-atom MCMD simulations with ACE potentials within reach.

The parallel MC algorithm of Sadigh \textit{et al.} \cite{sadigh_scalable_2012} is available in LAMMPS as \verb|fix sgcmc|. We first modified the ACE implementation in LAMMPS to work with this routine. It now evaluates $\Delta U$ for an identity-exchange move only from atoms within the interaction radius of the selected atom as
\begin{equation}
    \Delta U = \Delta U_i + \sum_{j \in \mathcal{N}_i} \Delta U_j
    \label{eq:atomic_energy_change}
\end{equation}
where $\Delta U_i$ is the atomic energy change for switching the species of atom $i$, and the sum runs over all atoms $j$ in the neighborhood $\mathcal{N}_i$ of atom $i$. Evaluating \cref{eq:atomic_energy_change} correctly for atoms close to subdomain boundaries requires adjusting the ghost-atom cutoff. Ghost atoms are atoms that LAMMPS communicates between subdomains to guarantee correct energies and forces. \Cref{fig:domain_boundary} shows the orange atom labeled R at the edge of a subdomain. The neighborhood of this atom contains ghost atoms such as the orange atom labeled G, and the neighborhoods of these ghost atoms contain further ghost atoms. The atomic energy change of atom G is evaluated correctly only if the ghost-atom region contains its entire neighborhood. For an arbitrary configuration, the ghost-atom region must therefore span at least twice the interaction radius $r_{\text{cut}}$ of the potential. This requirement is set in LAMMPS with the command \verb|comm_modify cutoff|.

Calculating $\Delta U$ from atomic energies makes the cost of an MC move independent of system size. In addition, \verb|fix sgcmc| carries out MC moves simultaneously in all computational subdomains, each of which corresponds to one MPI rank. MC wall times therefore decrease by a factor equal to the number of MPI ranks at a fixed number of CPUs. Parallel MC execution samples the SGC and VC-SGC ensembles correctly only if identity-exchange moves in different subdomains do not affect each other. This condition imposes a minimum subdomain size of four times $r_{\text{cut}}$ \cite{sadigh_scalable_2012}. For typical crystalline systems, a subdomain of this size contains 1000--5000 atoms, depending on the lattice parameter and $r_{\text{cut}}$, and a single MPI rank carries out the MD of all of them. This load is acceptable for empirical potentials owing to their low computational cost, but it slows down simulations with MLIPs. To circumvent this issue, we used shared-memory parallelization within each MPI rank. LAMMPS offers this through the \verb|KOKKOS| \cite{10.1145/3731599.3767498} package, and we added an \verb|OPENMP| variant of the ACE potential. Shared-memory evaluation of the ACE potential allows multiple OpenMP threads to be assigned to each MPI rank. As shown in \cref{fig:mpi_omp_parallelization}, MC is then parallelized at the level of MPI ranks, while MD is parallelized at the finer level of OpenMP threads. All simulations in this work used the \verb|OPENMP| variant.

\section{Free energy integration with MCMD} \label{sect:free_energy}
Larger simulation cells are useful both for modeling longer length scales and for computing free energies. Thermal fluctuations of the energy and composition decrease as $1 / \sqrt{N}$, where $N$ is the number of atoms in the simulation box, so larger systems reduce the statistical errors of free energy calculations. The hybrid parallelization scheme lowers the cost of these larger systems because MC wall times decrease at a fixed number of CPUs. Here, we describe how free energies are calculated from VC-SGC MCMD simulations.

Free energy integration requires the exchange chemical potential $\Delta \mu$ as a function of composition. The function $\Delta \mu (c)$ is obtained by equilibrating several simulation boxes at different target compositions with MCMD. The exchange chemical potential at each composition is calculated from the average composition according to \cite{sadigh_scalable_2012, sadigh_calculation_2012}
\begin{equation}
    \Delta \mu = -\frac{\partial g}{\partial c} = \phi + 2 N \kappa \langle c \rangle
    \label{eq:delta_mu_vc}
\end{equation}
where $\langle c\rangle$ is the ensemble average of the composition and $g$ is the Gibbs free energy per atom. The parameter $\phi$ is determined by the target composition $c_0$ and $\Delta \mu_{\text{guess}}$ through \cref{eq:phi_vc}. Integrating \cref{eq:delta_mu_vc} with respect to composition gives the free energy. 

Before integration, the exchange chemical potential data are converted into a smooth curve by spline interpolation. The ideal solution contribution to the exchange chemical potential diverges at $c = 0$ and $c = 1$, so we subtract it from the data before interpolating to ensure numerical stability. The interpolated exchange chemical potential is then
\begin{equation}
    \Delta \mu (c) = s(c) - k_B T \ln \left ( \frac{c}{1 - c} \right )
    \label{eq:spline_interp}
\end{equation}
where $s(c)$ is a cubic spline \cite{antillon_thermodynamical_2025}. Integrating \cref{eq:spline_interp} gives
\begin{equation}
    \begin{aligned}
        g(c) - g(0) = -\int_0^c s(c') dc' \\ + k_B T [ c \ln(c) + (1 - c) \ln(1 - c)]
    \end{aligned}
    \label{eq:free_energy_integral}
\end{equation}
where the second term is the contribution of the ideal entropy of mixing. \Cref{eq:free_energy_integral} gives the free energy only relative to $g(0)$. The reference free energy $g(0)$ must be calculated separately, for example with thermodynamic integration \cite{freitas_nonequilibrium_2016,menon_automated_2021}.

The efficiency of VC-SGC sampling depends on $\kappa$, the switching frequency between MC and MD, the number of attempted identity-exchange moves per MC routine, and $\Delta \mu_{\text{guess}}$. In line with previous studies \cite{sadigh_scalable_2012,sadigh_calculation_2012}, we found that values of $\kappa$ between $10^{-4}$ and $10^{-3}$ keep the average composition close to the target $c_0$ without reducing the MC acceptance probability. Smaller values are preferable as the number of MPI ranks and the system size grow. In this work, we switched between MC and MD every 10 time steps, and each MC routine attempted identity-exchange moves on 2.5\% of the atoms in the simulation box. The autocorrelation function of the composition time series indicated that samples were uncorrelated after 5--30 MC routines, depending on $\kappa$ and the number of MPI ranks. An accurate value of $\Delta \mu_{\text{guess}}$ maximizes the acceptance probability \cite{sadigh_calculation_2012} and improves the convergence of the free energies. We identified $\Delta \mu_{\text{guess}}$ in three stages. Two simulations at compositions close to 0 and 1 first bounded its range. Short MCMD trajectories with guess values in this range then provided a preliminary curve of the exchange chemical potential. The production runs took $\Delta \mu_{\text{guess}}$ from this curve. Additional computational details, including MD settings, are summarized in the supplementary material.

\section{Results} \label{sect:results}

\subsection{Scaling} \label{sect:scaling}

We assessed the hybrid scheme by measuring wall times for 10,000 time steps of VC-SGC MCMD of liquid Mg--Nd at 1100\,K with a previously developed ACE potential \cite{piersante_machine_2026}. The simulation box contained 17,248 atoms at a neodymium composition of $c_{\text{Nd}} = 0.1$ and had an edge length of approximately 78\,{\AA}. This length is roughly ten times $r_{\text{cut}}$. An MC routine was carried out every 100 time steps and attempted identity-exchange moves on 5\% of the atoms. \Cref{fig:scaling} reports the total wall times relative to the simulation with one MPI rank and 16 OpenMP threads, and breaks them down into MC and MD contributions. The first four configurations had a fixed total of 16 CPUs. Across these, the MC wall time decreased by a factor equal to the number of MPI ranks, while the MD wall time remained constant. Redistributing the same CPUs from threads to ranks therefore reduced the total wall time. In the last three configurations, increasing the number of threads per rank at eight ranks left the MC wall time unchanged and reduced the MD wall time. For comparison, a canonical MCMD simulation with the standard \verb|fix atom/swap| routine in LAMMPS on the same 16 CPUs took $\approx$6 times longer than the reference configuration.

The supplementary material compares the \verb|OPENMP| and \verb|KOKKOS| variants of the ACE potential. For standard MD of liquid Mg, the \verb|OPENMP| variant was faster, and a hybrid MPI/OpenMP configuration gave the shortest wall time. For the MCMD simulation of liquid Mg--Nd, the \verb|KOKKOS| variant was faster but required more memory. The relative performance of the two variants therefore depends on the simulation type and the parallel configuration, and it is likely to depend on the computing platform as well.

\begin{figure*}[!ht]
    \centering
    \subfloat[][\label{fig:HfTi_snapshots}]{\includegraphics[width=0.9\linewidth]{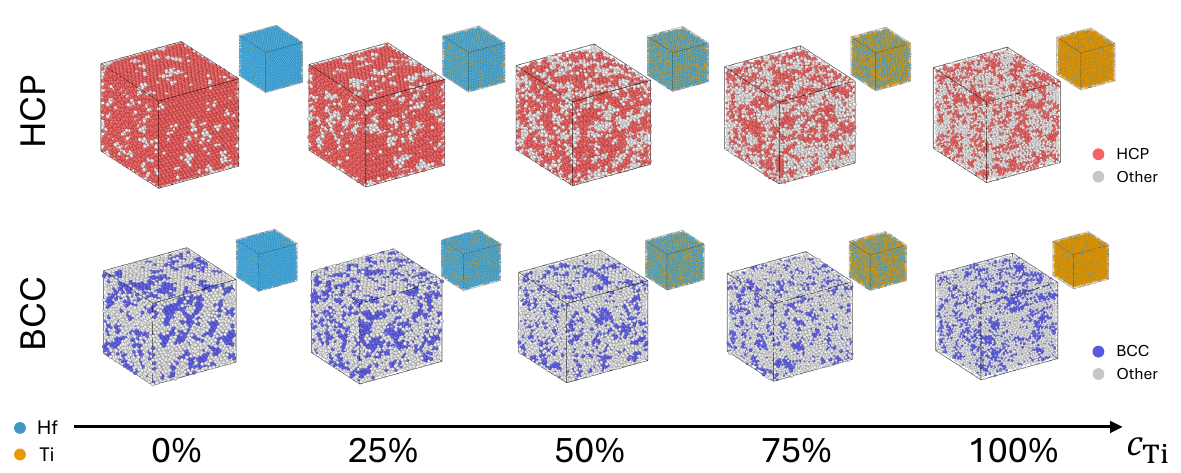}} \\
    \subfloat[][\label{fig:HfTi_lattice_params}]{\includegraphics[height=0.25\linewidth]{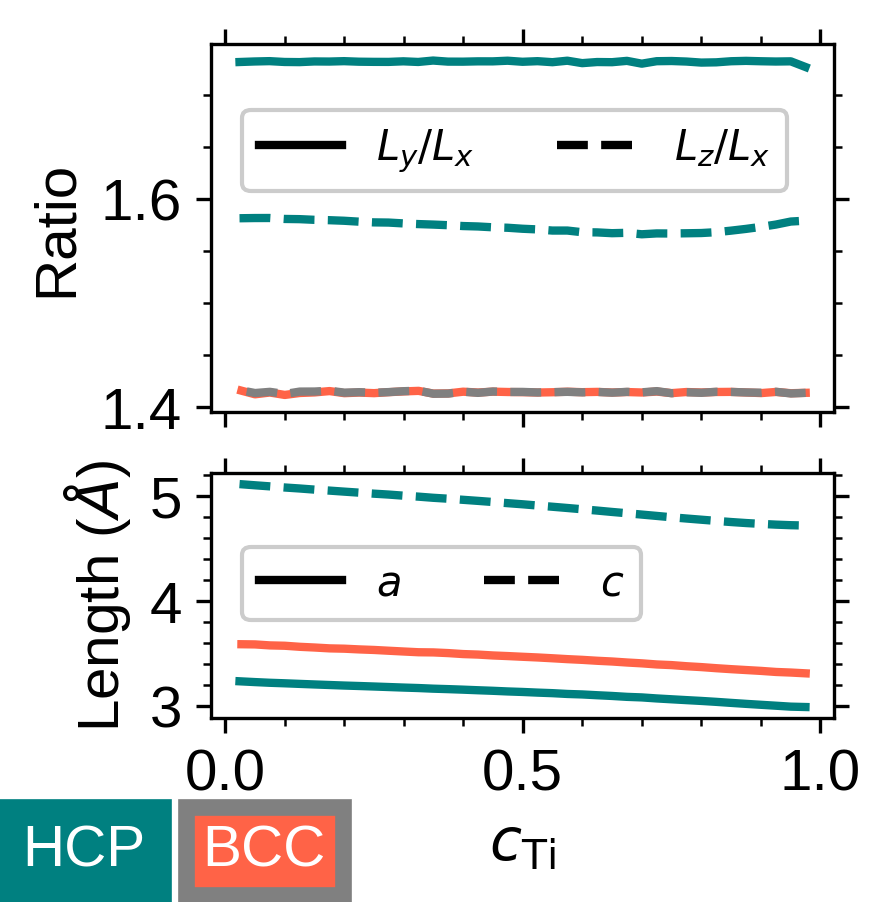}}
    \subfloat[][\label{fig:HfTi_wall_times}]{\includegraphics[height=0.25\linewidth]{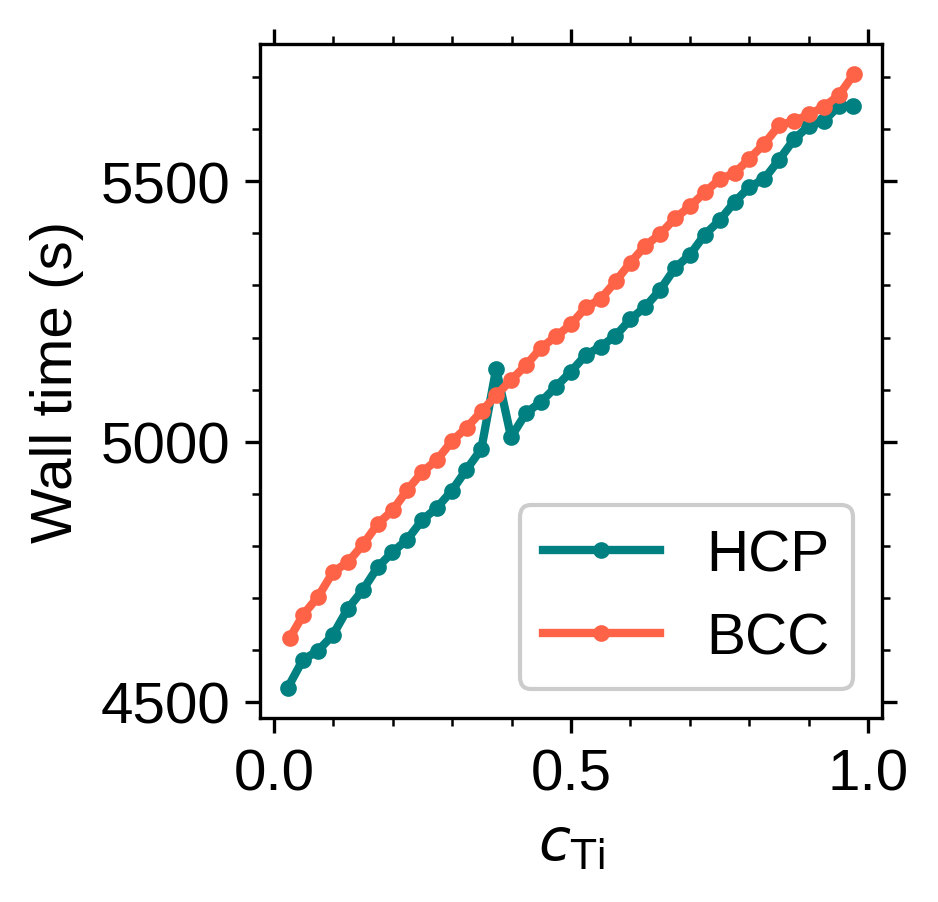}}
    \subfloat[][\label{fig:HfTi_chem_pot}]{\includegraphics[height=0.25\linewidth]{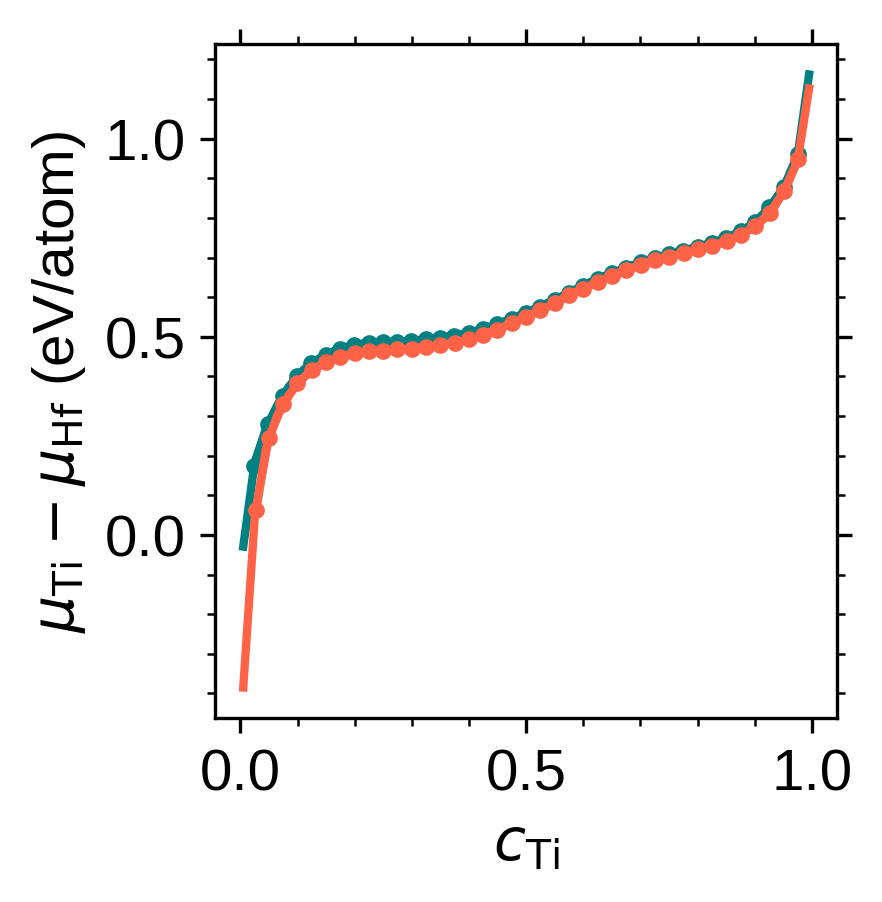}}
    \subfloat[][\label{fig:HfTi_free_energy}]{\includegraphics[height=0.25\linewidth]{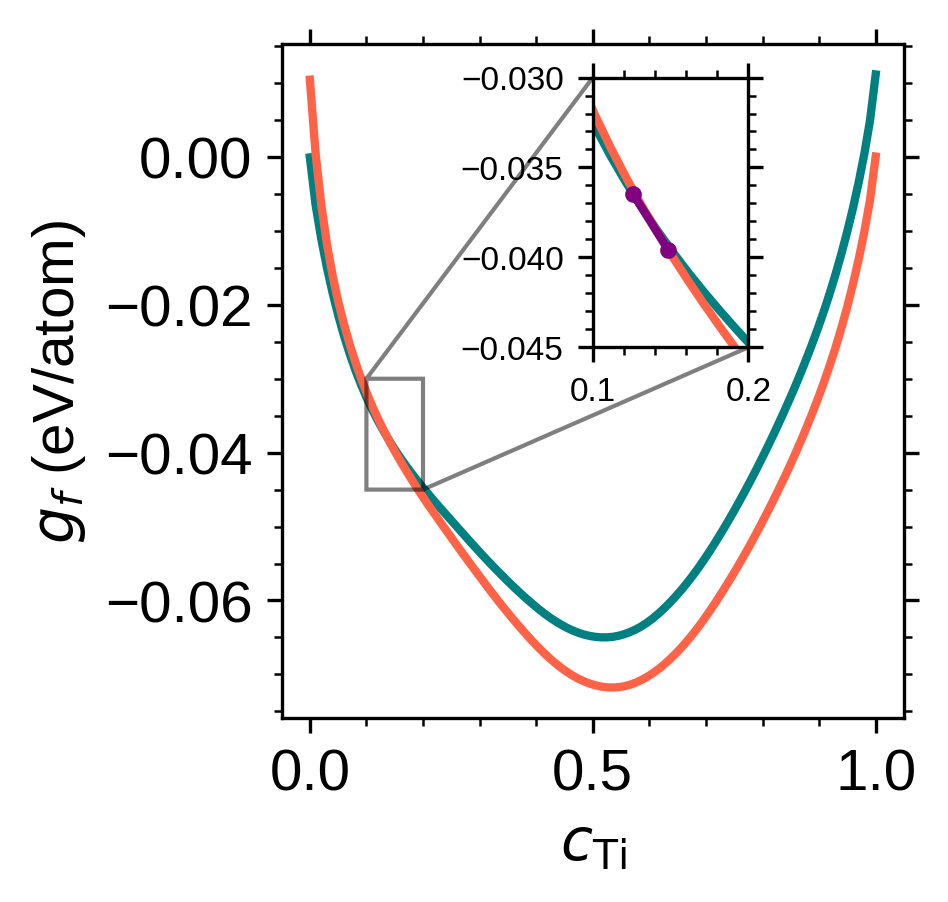}}
    \caption{(a) Composition sweeps for the HCP (top) and BCC (bottom) phases of Hf--Ti at 1500\,K. Atoms in the large simulation boxes are colored by the common neighbor analysis in OVITO, while the small boxes show the chemical species. Thermal fluctuations cause many atoms to be classified as ``Other''. (b) Ratios of the simulation-box lengths (top) and lattice parameters (bottom) as a function of the titanium composition $c_{\text{Ti}}$. (c) Wall time of the simulation at each composition on 72 CPUs with eight MPI ranks. (d) Exchange chemical potential as a function of titanium composition. Dots mark the MCMD averages and solid lines the interpolation with \cref{eq:spline_interp}. (e) Free energies of the HCP and BCC phases relative to HCP Hf and BCC Ti at 1500\,K as a function of titanium composition. The inset shows the common tangent construction. %
    }
\end{figure*}

\subsection{Hf--Ti} \label{sect:HfTi}
MCMD in the VC-SGC ensemble gives free energies that account for chemical ordering and lattice vibrations on the same footing. This treatment is relevant for dynamically stabilized phases because local chemistry is expected to couple to anharmonic lattice vibrations. We applied it to the refractory Hf--Ti system, in which the BCC phase is dynamically stabilized at high temperature. The goal of this example is to demonstrate the method and its implementation in LAMMPS. The HCP-to-BCC transition temperature is 2016\,K in hafnium and 1155\,K in titanium \cite{DINSDALE1991317}, so the alloy transforms from HCP to BCC with increasing titanium content at intermediate temperatures. The free energies from MCMD give the composition range of the HCP--BCC two-phase region.

We calculated free energies at 1500\,K with the procedure of \cref{sect:free_energy}. The MCMD simulations used a previously developed ACE potential \cite{liyanage2026crackingcasefluctuationsenhance} and a 17,248-atom orthorhombic simulation box, and they ran on 72 CPUs distributed over eight MPI ranks. With eight ranks, the MC wall time was close to that of a cell one eighth the size. As illustrated in \cref{fig:HfTi_snapshots}, we carried out two composition sweeps starting from HCP and BCC hafnium. The target titanium composition was increased in increments of 0.025, and each composition was equilibrated for 5000 time steps and sampled for a further 5000.

Large anharmonic fluctuations at 1500\,K distort the local coordination, so the common neighbor analysis classifies a large fraction of the atoms as neither HCP nor BCC (\cref{fig:HfTi_snapshots}). We therefore identified the crystal structure from the shape of the simulation box. The box is built from an orthorhombic cell that accommodates the Burgers transformation between BCC and HCP \cite{raju_natarajan_toward_2019}. In this cell, the edge lengths of an HCP crystal are in the ratio $1 : \sqrt{3} : c/a$, while those of a cubic BCC crystal are in the ratio $1 : \sqrt{2} : \sqrt{2}$. The top panel of \cref{fig:HfTi_lattice_params} shows that $L_y / L_x$ remained close to $\sqrt{3}$ for HCP and that both ratios remained close to $\sqrt{2}$ for BCC, so both cells kept their crystal structure throughout the sweeps. The lattice parameters of both phases decreased smoothly with titanium content (bottom panel of \cref{fig:HfTi_lattice_params}). The wall time increased with titanium content (\cref{fig:HfTi_wall_times}) because each atom has more neighbors within the fixed $r_{\text{cut}}$ in the denser cells. Equivalent canonical MCMD simulations with the standard \verb|fix atom/swap| routine of LAMMPS would have taken $\approx$18 times longer.

Block averaging \cite{frenkel_understanding_2023} gave average statistical errors of $\approx 2.2 \times 10^{-5}$ in the composition and 0.75\,meV/atom in the exchange chemical potential. \Cref{fig:HfTi_chem_pot} shows the exchange chemical potential of both phases as a function of titanium composition, together with the interpolation of \cref{eq:spline_interp}. Integrating these curves with \cref{eq:free_energy_integral} gave the free energies in \cref{fig:HfTi_free_energy}. The free energy $g_f$ is the Gibbs free energy of the alloy referenced to the free energies of HCP hafnium and BCC titanium at 1500\,K. The reference free energies of HCP and BCC hafnium were calculated with equilibrium Frenkel--Ladd thermodynamic integration \cite{frenkel_new_1984}. The common tangent construction in the inset of \cref{fig:HfTi_free_energy} places the HCP--BCC two-phase region between $c_{\text{Ti}} = 0.126$ and 0.149. The HCP phase is stable below this range and the BCC phase above it.

\begin{figure*}[!ht]
    \centering
    \includegraphics[width=0.9\linewidth]{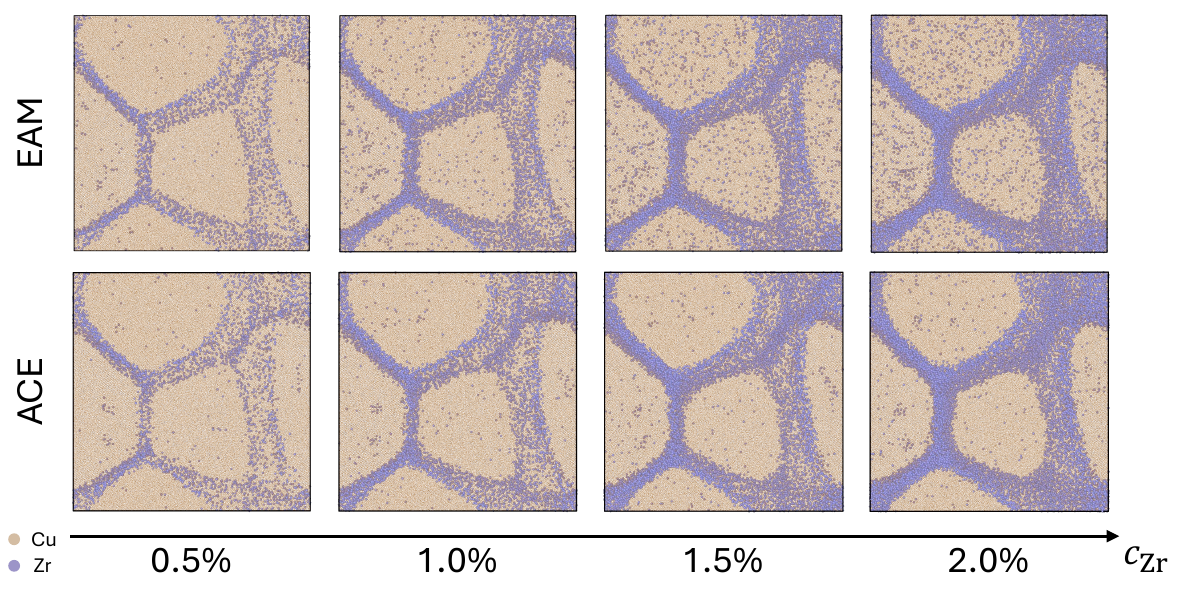}
    \caption{Zirconium segregation in nanocrystalline copper at 1000\,K as predicted by the EAM (top) and ACE (bottom) potentials. The global zirconium composition $c_{\text{Zr}}$ increases from 0.5\% to 2.0\% from left to right. Copper atoms are shown in brown and zirconium atoms in purple.}
    \label{fig:CuZr_snapshots}
\end{figure*}

\begin{figure}
    \centering
    \subfloat[][\label{fig:CuZr_grain_fraction}]{\includegraphics[height=0.5\linewidth]{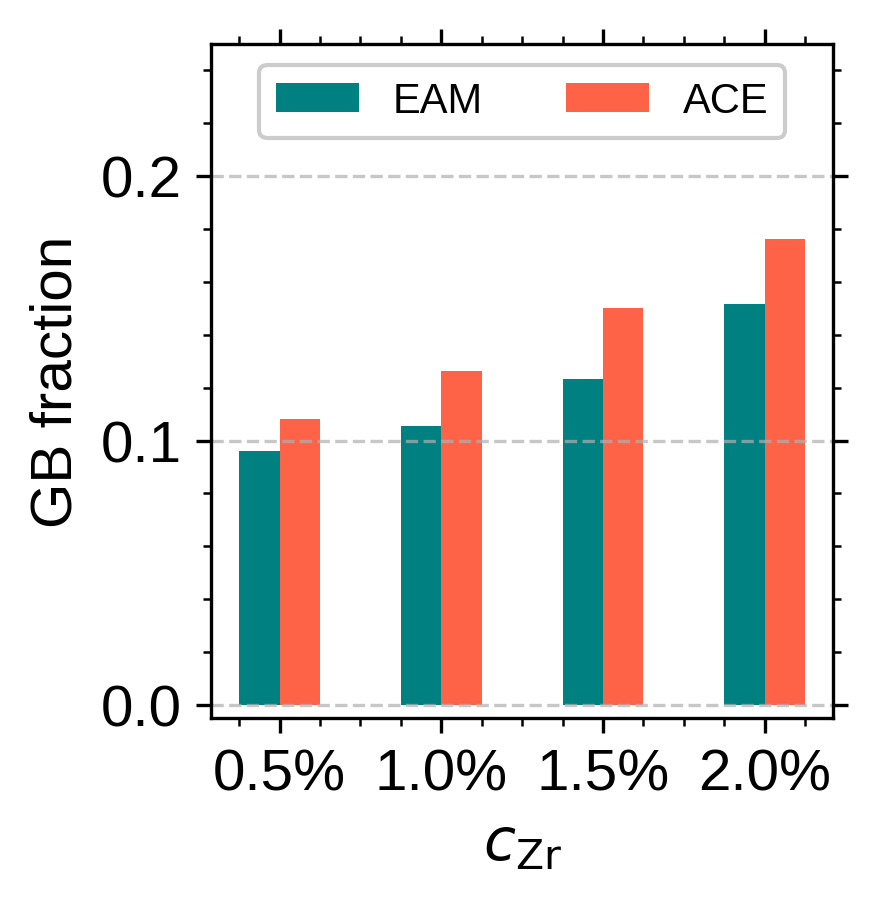}}
    \subfloat[][\label{fig:CuZr_gb_bulk_comp}]{\includegraphics[height=0.5\linewidth]{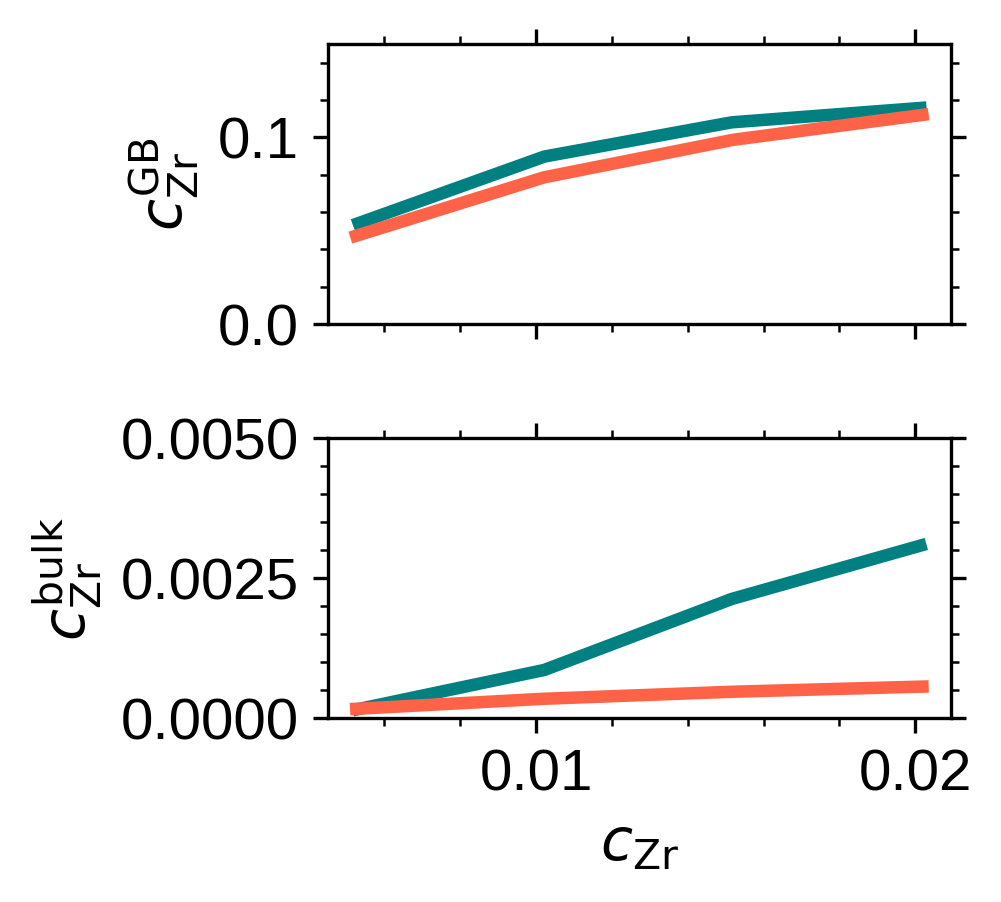}}
    \caption{(a) Grain boundary (GB) fraction and (b) zirconium composition of the grain boundaries (top) and of the bulk (bottom) as a function of the global zirconium composition $c_{\text{Zr}}$. Atoms with FCC coordination are counted as bulk sites and all remaining atoms as GB sites. %
    }
\end{figure}

\subsection{Cu--Zr} \label{sect:CuZr}
The hybrid scheme enables MCMD simulations of crystalline defects such as dislocations and grain boundaries with ACE potentials at large system sizes. As a demonstration, we compared grain boundary segregation in dilute Cu--Zr alloys as predicted by the EAM potential of Mendelev \textit{et al.} \cite{mendelev_development_2019} and by the ACE potential of Leimeroth \textit{et al.} \cite{leimeroth_general_2024}. Without the hybrid scheme, a direct comparison at this system size would be impractical because of the cost of MCMD with the ACE potential.

Cu--Zr is a metallic-glass-forming alloy \cite{mendelev_development_2019,leimeroth_general_2024} and a common model system for grain boundary complexions \cite{pan_effect_2016,nawi_contribution_2026}. Zirconium is almost insoluble in copper and therefore segregates strongly to grain boundaries. Simulations of complexions require large systems, so many studies have relied on empirical potentials \cite{pan_effect_2016,mendelev_development_2019,nawi_contribution_2026,sansoz_disordered_2026}, and how their predictions compare with those of MLIPs is not known.

Following Nawi \textit{et al.} \cite{nawi_contribution_2026}, we used Atomsk \cite{HIREL2015212} to create a cubic nanocrystalline copper sample with an edge length of 200\,{\AA} (670,000 atoms). The sample was equilibrated for 200\,ps at 1000\,K with each potential. We then ran VC-SGC MCMD at 1000\,K for 200,000 time steps at target compositions from $c_{\text{Zr}} = 0.005$ to 0.02 in increments of 0.005, with the MC settings of \cref{sect:free_energy}. The compositions were simulated sequentially, and each run started from the final configuration of the previous one. We set $\Delta \mu_{\text{guess}}$ to 3.872\,eV/atom for the EAM potential and 3.182\,eV/atom for the ACE potential. These values were determined from VC-SGC MCMD simulations of $11 \times 11 \times 11$ FCC supercells at a target zirconium composition of 0.02.

Configurations were collected every 5000 time steps over the last quarter of each trajectory, and thermal noise was removed by energy minimization. The common neighbor analysis in OVITO then classified atoms with FCC coordination as bulk sites and all remaining atoms as grain boundary (GB) sites. The GB fraction is $1 - N^{\text{FCC}}/N$, where $N^{\text{FCC}}$ is the number of FCC-coordinated atoms. We denote the number of zirconium atoms by $N_{\text{Zr}}$ and the number of FCC-coordinated zirconium atoms by $N_{\text{Zr}}^{\text{FCC}}$. The zirconium compositions of the bulk and of the grain boundaries are then $c_{\text{Zr}}^{\text{bulk}} = N_{\text{Zr}}^{\text{FCC}} / N^{\text{FCC}}$ and $c_{\text{Zr}}^{\text{GB}} = (N_{\text{Zr}} - N_{\text{Zr}}^{\text{FCC}}) /(N - N^{\text{FCC}})$.

\Cref{fig:CuZr_snapshots} shows the equilibrated structures obtained with the two potentials. Zirconium segregated strongly to the grain boundaries with both potentials, and the structures at the lowest global zirconium composition are visually similar. As the zirconium composition increased, the EAM potential placed more zirconium atoms inside the grains than the ACE potential. The GB fraction was higher with the ACE potential at every composition (\cref{fig:CuZr_grain_fraction}). The grain boundary compositions of the two potentials were close, with slightly lower values for the ACE potential (\cref{fig:CuZr_gb_bulk_comp}). The bulk compositions differed more strongly. For the EAM potential, $c_{\text{Zr}}^{\text{bulk}}$ increased steadily with the global composition, in line with a previous study \cite{pan_effect_2016}, while for the ACE potential it remained nearly constant and close to zero. This difference can affect the mechanical response of the grains through solid-solution strengthening. The two potentials therefore agree on the grain boundary composition but differ quantitatively in the amount of zirconium dissolved in the grains.

\section{Discussion}

Large-scale MCMD with MLIPs has applications beyond the two examples of this work. The coupling between local chemistry and lattice vibrations treated in the Hf--Ti example also governs the formation of intermetallic phases in multicomponent alloys, where chemical ordering is coupled to structural transformations \cite{natarajan_connecting_2018,kolli_six_2021}. Direct simulations of these coupled transformations therefore require accurate MLIPs and large simulation cells. In the Cu--Zr example, large simulation cells were applied to segregation at grain boundaries. The same combination of free energy methods and large simulation cells can provide access to finite-temperature surface and interface free energies in the presence of chemical disorder. Examples include the diffuse interfaces that form during nucleation or spinodal decomposition at high temperature \cite{clouet_nucleation_2004,lepinoux__interfacial_2005,stechauner_thermo-kinetic_2015}. The structure of these interfaces is required to interpret experimental observations and is an input to mesoscale models based on classical nucleation theory, cluster dynamics or phase field methods \cite{deschamps_precipitation_2021}. The VC-SGC ensemble can also be used to calculate phase diagrams, although its advantages and disadvantages relative to other methods for estimating phase boundaries or free energies, such as alchemical transformations \cite{menon_computing_2026}, remain to be assessed.

Despite the acceleration provided by the hybrid scheme, the computational cost of the MLIP remains the main factor in the total simulation time. It sets the system size and trajectory length that are practical. Leimeroth \textit{et al.} \cite{leimeroth_general_2024} limited their ACE potential to 514 descriptors with large-scale simulations of the mechanics of metallic glasses in mind, and the MCMD simulations of nanocrystalline Cu--Zr benefited from this low cost. Workflows that tune the hyperparameters of an MLIP to the accuracy and speed that a given modeling task requires remain to be developed.

\section{Conclusion}

We introduced a hybrid MPI/OpenMP scheme for parallel MCMD simulations with ACE potentials in LAMMPS. MC is parallelized over MPI ranks by \verb|fix sgcmc|, while MD is parallelized at the finer level of OpenMP threads. We reviewed the SGC and VC-SGC ensembles and their parallel MC implementation, and outlined the procedure for calculating free energies from parallel VC-SGC simulations. Benchmarks for liquid Mg--Nd showed that, at a fixed number of CPUs and atoms, redistributing CPUs from threads to ranks reduces the MC wall time by a factor equal to the number of ranks. We applied the scheme to the HCP and BCC free energies of the strongly anharmonic Hf--Ti system and to zirconium segregation in nanocrystalline copper. For Cu--Zr, the EAM potential of Mendelev \textit{et al.} \cite{mendelev_development_2019} and the ACE potential of Leimeroth \textit{et al.} \cite{leimeroth_general_2024} agreed on the grain boundary composition but differed in the amount of zirconium dissolved in the grains. The choice between an empirical potential and an MLIP should therefore depend on the accuracy that the intended application requires. The methods and implementation of this study make large-scale MCMD simulations with MLIPs practical and enable rigorous modeling of the interactions between chemical ordering, lattice vibrations and crystalline defects.

\section{Supplementary material}
The supplementary material contains a strong scaling analysis of \verb|pair style pace/omp| for standard MD, a comparison between the \verb|OPENMP| and \verb|KOKKOS| variants for MCMD, a comparison between free energies obtained with the parallel VC-SGC and the SGC ensemble, a study of the effect of $\kappa$ on the acceptance probability of VC-SGC MCMD, an analysis of the composition autocorrelation function as a function of $\kappa$ and the number of MPI ranks, and a summary of computational details (MD settings, $\kappa$ values and thermodynamic integration settings).

\section{Acknowledgments}
This research was supported by the Swiss National Science Foundation (Grant No. 215178). ARN is grateful for partial support through the framework of the EUROfusion Consortium, partially funded by the European Union via the Euratom Research and Training Programme (grant agreement no. 101052200---EUROfusion). Views and opinions expressed are however those of the author(s) only and do not necessarily reflect those of the European Union or the European Commission. Neither the European Union nor the European Commission can be held responsible for them.

\section{Author declarations}

\subsection{Conflict of interest}
The authors have no conflicts to disclose. 

\subsection{Author contributions}
\textbf{Lorenzo Piersante}: Conceptualization (lead); Formal analysis; Investigation; Methodology; Software; Visualization; Writing/Original Draft Preparation.
\textbf{Anirudh Raju Natarajan}: Conceptualization (supporting); Funding Acquisition; Resources; Supervision; Writing/Review \& Editing.

\section{Data availability}
The data that support the findings of this study are available from the corresponding author upon reasonable request. The shared-memory variant of the ACE potential is available as \verb|pair style pace/omp| in the \verb|OPENMP| package of the official LAMMPS release. The local evaluation of ACE potentials in \verb|fix sgcmc| is enabled with the flag \verb|atomic/energy yes| and is currently available in a fork of LAMMPS at \url{https://github.com/LorenzoPiersante/lammps}.

\bibliographystyle{unsrt}
\bibliography{bibliography}

\end{document}


\title{Supplementary material: Large-scale Monte Carlo/Molecular Dynamics simulations with Atomic Cluster Expansion potentials in LAMMPS}

\author{Lorenzo Piersante}
\email{lorenzo.piersante@epfl.ch}
\affiliation{Laboratory of Materials Design and Simulation (MADES), Institute of Materials, \'{E}cole Polytechnique F\'{e}d\'{e}rale de Lausanne}

\author{Anirudh Raju Natarajan}
\email{anirudh.natarajan@epfl.ch}
\affiliation{Laboratory of Materials Design and Simulation (MADES), Institute of Materials, \'{E}cole Polytechnique F\'{e}d\'{e}rale de Lausanne}

\date{\today}

\maketitle

\section{Strong scaling of the OPENMP ACE compute}

\begin{figure*}[!ht]
    \centering
    \subfloat[][\label{fig:mpi_openmp_comparison}]{\includegraphics[height=0.3\linewidth]{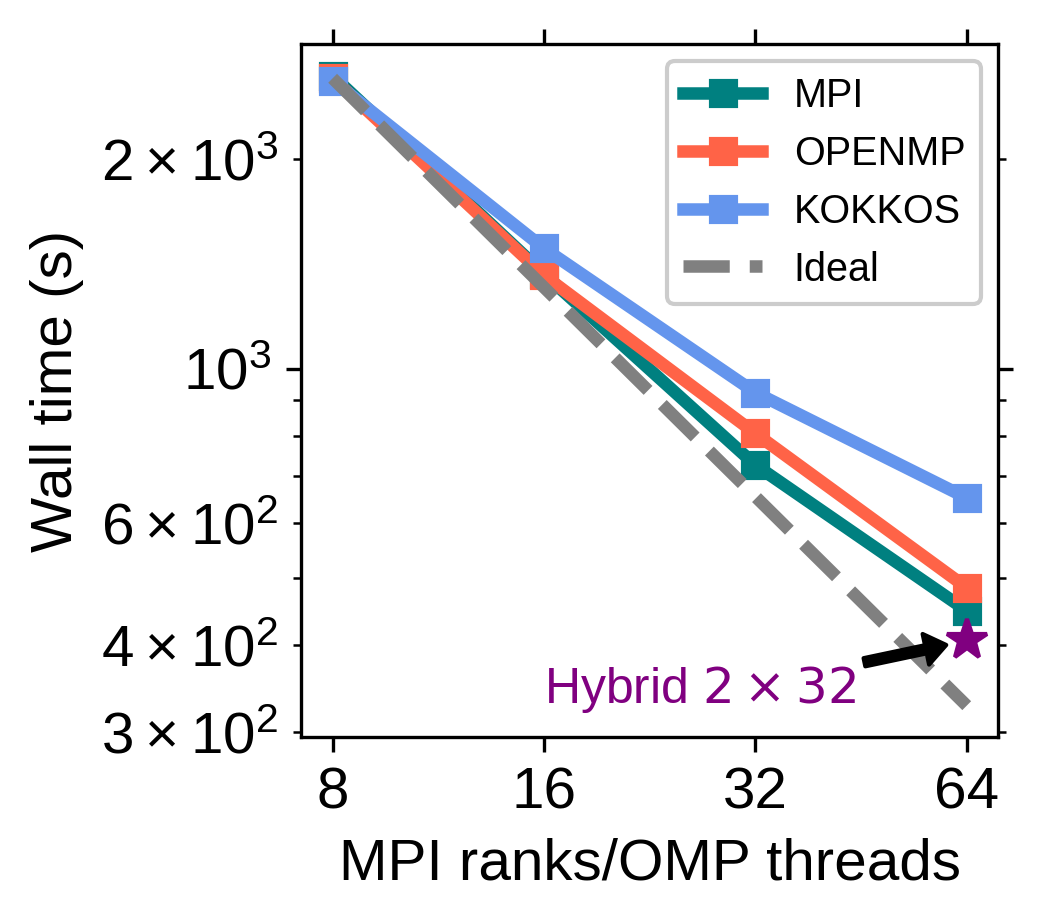}}
    \subfloat[][\label{fig:mcmd_openmp_kokkos_comparison}]{\includegraphics[height=0.3\linewidth]{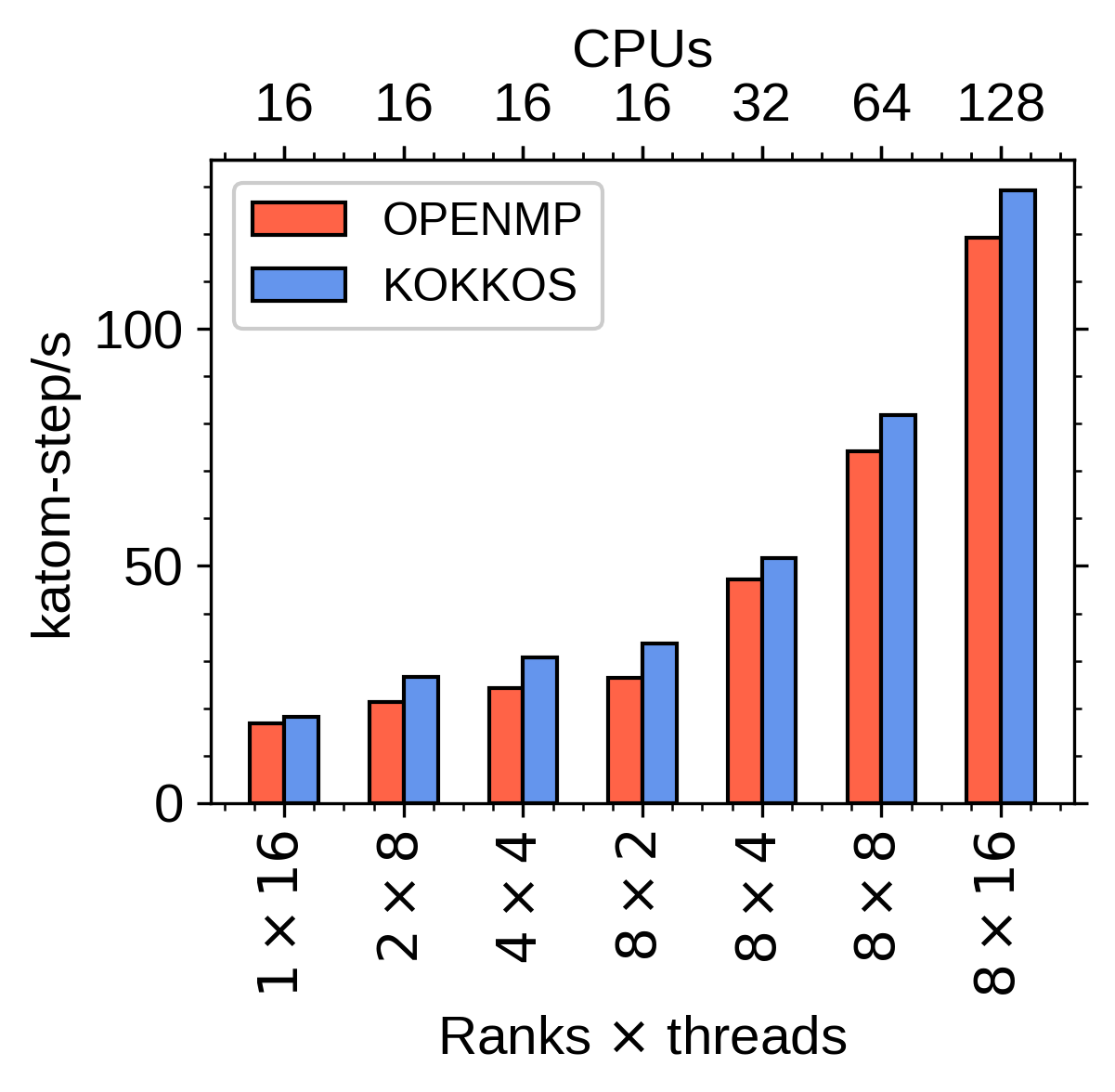}}
    \caption{(a) Strong scaling study for plain molecular dynamics with the different variants of the ACE potential. (b) Comparison of MCMD efficiency between the OPENMP and KOKKOS variants.}
\end{figure*}

We investigated the strong scaling of the \verb|OPENMP| ACE compute (\verb|pair style pace/omp|) we implemented and compared it to the standard MPI implementation (\verb|pair style pace|) and the recently developed \verb|KOKKOS| multithreaded version (\verb|pair style pace/kk|) \cite{10.1145/3731599.3767498}. Strong scaling measures the execution speedup as the number of CPUs increases. We measured the wall time for running 20,000 molecular dynamics steps with a previously developed Mg--Nd potential \cite{piersante_machine_2026} in a 2156-atom simulation box of liquid Mg. \Cref{fig:mpi_openmp_comparison} reports the outcomes of our test. One CPU is attached to each MPI rank or OpenMP thread. The dashed line indicates ideal scaling calculated relative to the 8-CPU MPI simulation. The \verb|OPENMP| variant performs identically to MPI up to 16 CPUs, beyond which its performance deteriorates slightly. The \verb|KOKKOS| version loses efficiency already at 16 threads, the gap relative to MPI widening as the amount of resources increases. One can regain some of the lost efficiency via hybrid MPI/OpenMP, i.e., by attaching multiple threads to an MPI task. The plot shows an example of execution with 2 ranks and 32 threads using the \verb|OPENMP| compute. The wall time reduces by $\approx$10\% relative to pure MPI execution. When we attempted the same parallel configuration with the \verb|KOKKOS| variant, the simulation remained $\approx$12\% more expensive than pure MPI. Trying out different hybrid configurations especially benefits large-scale simulations, where gains of even a few percent are noticeable.

\textbf{ALT TEXT}
Figure S1a shows the variation of the wall times (in seconds) for plain molecular dynamics of liquid magnesium using the MPI, \verb|OPENMP|, and \verb|KOKKOS| ACE variants available in LAMMPS as the number of MPI ranks or OpenMP threads increases from 8 to 64. One CPU is attached to each rank or thread. The MPI variant is consistently the fastest, followed by \verb|OPENMP|, and then \verb|KOKKOS|. The star marker at 64 CPUs indicates that the hybrid configuration of 2 ranks and 32 threads with the \verb|OPENMP| ACE variant brings the wall time below that of pure MPI execution. Figure S1b is a bar chart of MCMD throughput (molecular dynamics steps per second of compute time for a 1000-atom system) when employing hybrid MPI/OpenMP parallelization with either the \verb|OPENMP| or the \verb|KOKKOS| package. The system considered is Mg--Nd at 0.1 neodymium concentration and 1100\,K. The bottom $x$-axis reports the hybrid MPI/OpenMP configuration (ranks $\times$ threads), while the top $x$-axis records the number of CPUs, ranging from 16 to 128. Multithreading with \verb|KOKKOS| is systematically more efficient. However, the performance gap narrows as the number of threads grows.

\section{Comparison of MCMD efficiency between OPENMP and KOKKOS accelerator variants}
We repeated the same scaling analysis of \cref{main-sect:scaling} for MCMD in liquid Mg--Nd using the \verb|KOKKOS| ACE accelerator variant. \Cref{fig:mcmd_openmp_kokkos_comparison} compares MCMD wall times, reported in katoms-step per second, for hybrid MPI/OpenMP parallelization employing either the \verb|OPENMP| or the \verb|KOKKOS| package. This quantity measures the number of time steps completed in one second of compute time for a system of 1000 atoms. On average, \verb|KOKKOS| showed a better performance relative to \verb|OPENMP|, with a speedup of around $17\pm 9 \%$. As the number of threads increases, the efficiency gain decreases from 28\% with 2 threads to 8\% with 16 threads. However, this performance gain comes with a higher memory footprint compared to the \verb|OPENMP| package. \verb|OPENMP| multithreading has memory requirements identical to, if not lower than, those of pure MPI execution. \verb|KOKKOS| multithreading ensures thread-safety by copying instances of the system onto each thread, which demands more memory. In general, the scaling results for plain molecular dynamics (\cref{fig:mpi_openmp_comparison}) and MCMD (\cref{fig:mcmd_openmp_kokkos_comparison}) indicate that the performance of the two packages depends on simulation type, parallelization setup, and underlying computing platform. Users should carry out appropriate benchmarks before production runs.

\section{Comparison between SGC and VC-SGC ensembles for free energy integration}
\begin{figure*}[!ht]
    \includegraphics[height=0.3\linewidth]{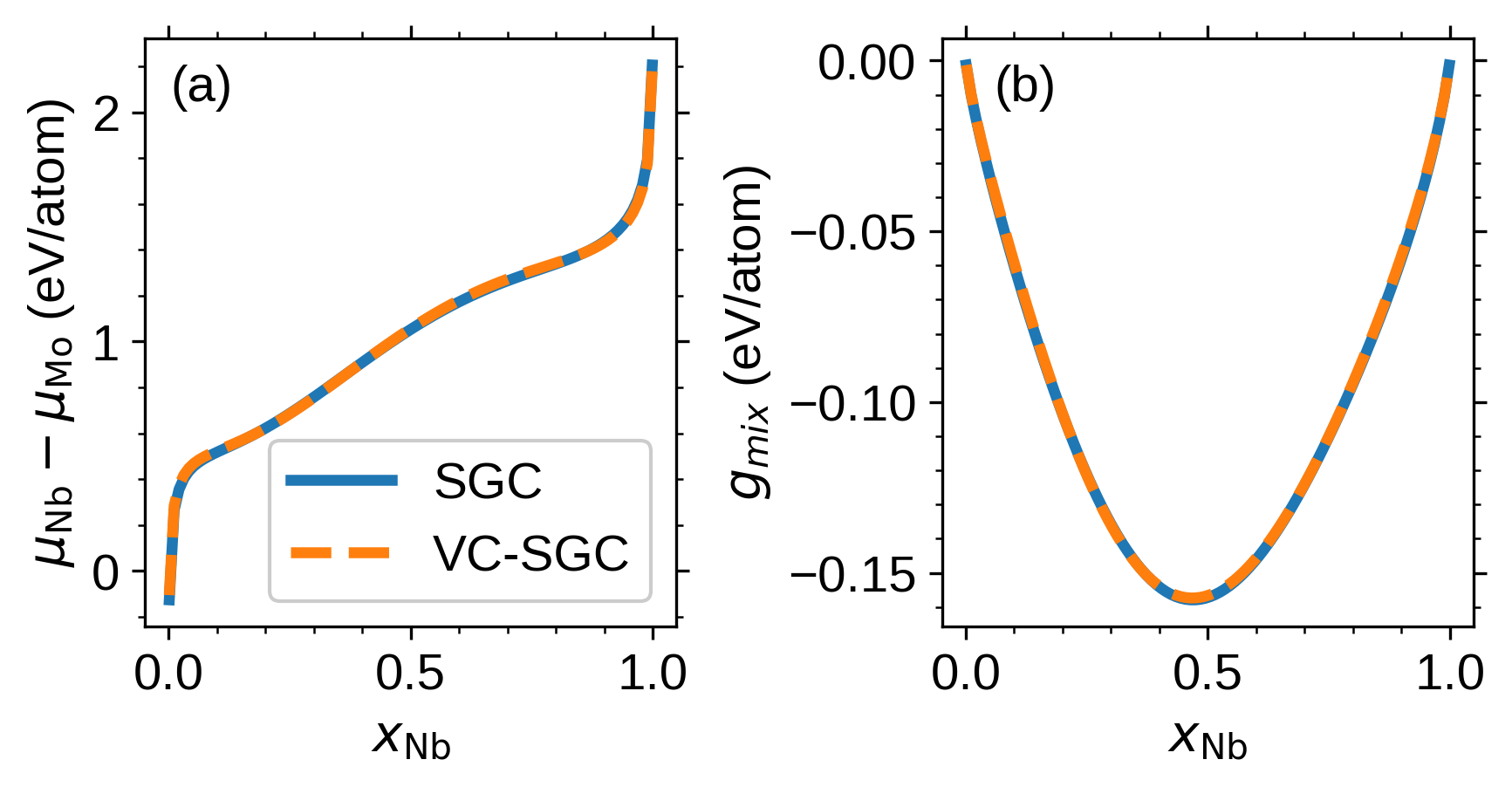}
    \caption{(a) Chemical potential as a function of niobium composition calculated with the SGC and VC-SGC ensembles. (b) Mixing free energy derived by integrating the chemical potential.}
    \label{fig:MoNb_free_energy}
\end{figure*}

To ensure the correctness of our free energy integration procedure, we verified that the standard SGC ensemble and the parallel VC-SGC ensemble produced the same results. We used Mo--Nb as a test system because at high temperature this alloy remains disordered for all compositions. We ran MCMD at 2000\,K within the two ensembles with a previously developed Mo--Nb potential \cite{liyanage2026crackingcasefluctuationsenhance}. We followed the sampling procedure outlined in the main text with $\kappa = 10^{-3}$ and, for simplicity, we used the chemical potential derived from the SGC ensemble as guess chemical potential for the VC-SGC ensemble. The results of the test are shown in \cref{fig:MoNb_free_energy}. The chemical potentials derived with the two methods agree (\cref{fig:MoNb_free_energy}a), and consequently the mixing free energies also match (\cref{fig:MoNb_free_energy}b).

\textbf{ALT TEXT}
Figure S2a shows the chemical potential (in eV/atom) against niobium composition curves for Mo--Nb at 2000\,K calculated using either the SGC or the VC-SGC ensembles. The two curves overlap almost exactly. Figure S2b is the mixing free energy (in eV/atom) against niobium composition derived by integrating the two chemical potential curves. Also in this case, the two curves match almost exactly.

\section{Variance-constraint parameter}

Selecting a reasonable variance-constraint parameter $\kappa$ is important to efficiently sample the VC-SGC ensemble. A small $\kappa$ might be insufficient to attain the desired equilibrium concentration, whereas a large $\kappa$ could suppress the composition fluctuations needed to explore microstates adequately. We examined the influence of $\kappa$ on VC-SGC MCMD simulations of Hf--Ti at 1500\,K with a target titanium concentration of 0.5\,.

\begin{figure}[!ht]
    \centering
    \includegraphics[height=0.3\linewidth]{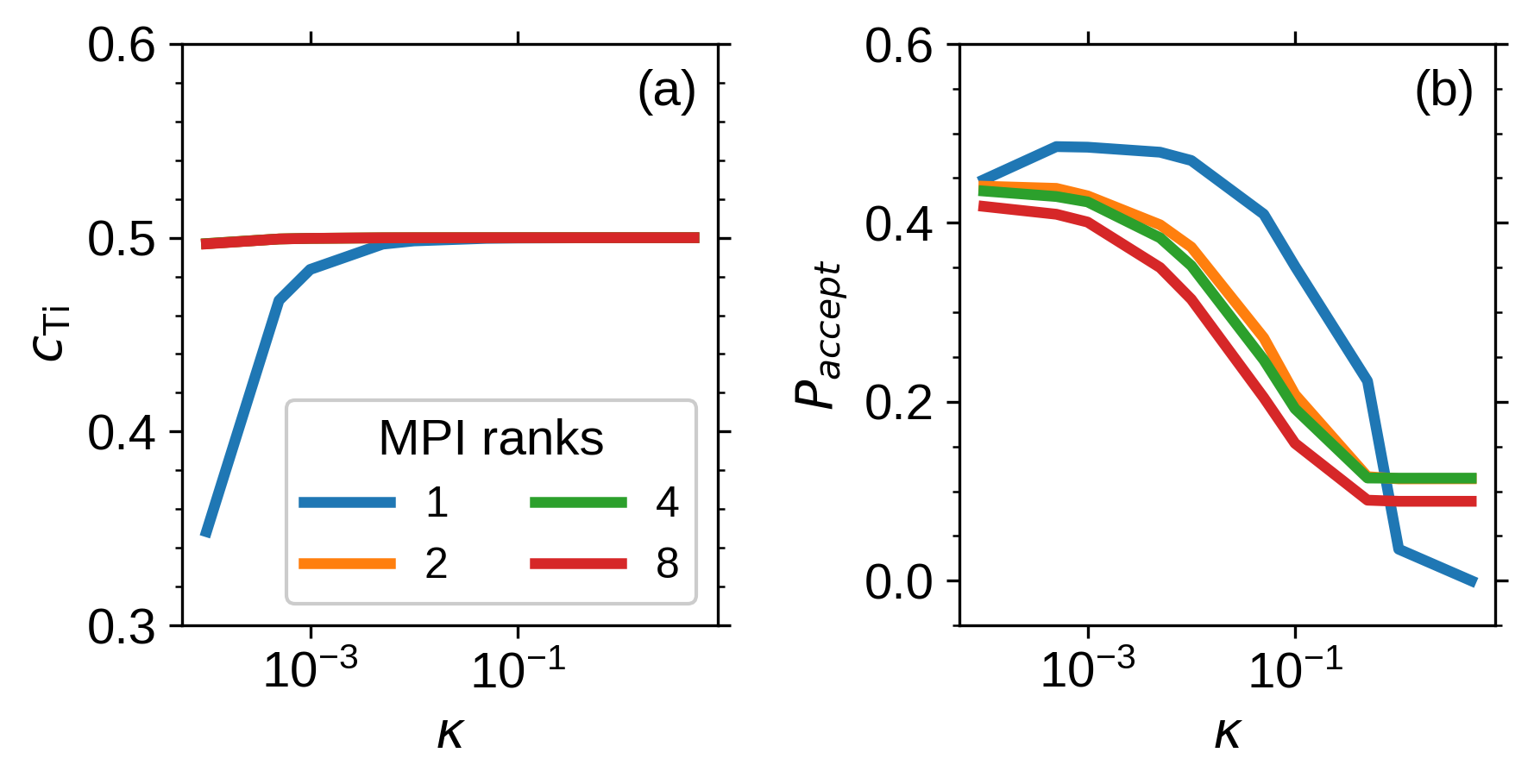}
    \caption{(a) Variation of the titanium concentration with variance-constraint parameter $\kappa$ and number of MPI ranks. (b) Change in acceptance probability with $\kappa$ and MPI ranks.}
    \label{fig:acceptance_kappa}
\end{figure}

We analyzed the effect of $\kappa$ and the number of MPI ranks (i.e. the number of computational subdomains across which MC is parallelized) on the average composition and MC acceptance probability. Using the same simulation settings as in \cref{main-sect:free_energy,main-sect:HfTi}, we changed the number of ranks from one (serial VC-SGC algorithm) to eight (parallel), while varying $\kappa$ at the same time.  \Cref{fig:acceptance_kappa} summarizes the results of the parameter study. \Cref{fig:acceptance_kappa}a shows that the serial algorithm deviates from the target composition for small $\kappa$ and needs a value of at least $10^{-3}$ to approach the target 0.5 composition. Employing parallel MC sets the average composition exactly. However, this reduces the maximum acceptance probability relative to the serial algorithm, as illustrated by \cref{fig:acceptance_kappa}b. Serial VC-SGC MC reaches a maximum acceptance probability of $\approx$0.48, after which it drops with increasing $\kappa$. The curves for two, four, and eight MPI ranks begin at 0.44 and then monotonically decline, their shape being similar. As the number of ranks increases, the maximum acceptance probability systematically shifts down, but the reduction is small. 

\begin{figure}[!ht]
    \centering
    \includegraphics[height=0.22\linewidth]{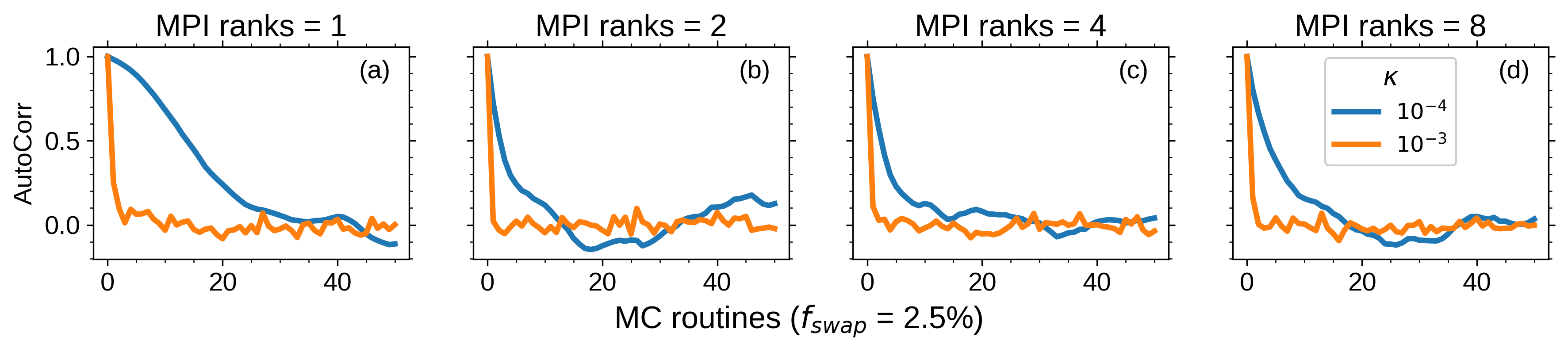}
    \caption{Autocorrelation  functions for $\kappa = 10^{-4}$ and $10^{-3}$ as the number of MPI ranks is increased from one (a) through eight (d).}
    \label{fig:autocorrelation_kappa}
\end{figure}

We further processed the composition timeseries to extract autocorrelation functions. These curves express how long it takes for two samples to become uncorrelated, which impacts the efficiency of MC sampling. \Cref{fig:autocorrelation_kappa} presents the autocorrelation functions for one, two, four, and eight MPI ranks for $\kappa = 10^{-4}$ and $10^{-3}$. The $x$-axis on the plots counts the number of MC routines rather than the number of molecular dynamics steps, because the latter leave the composition unchanged. \Cref{fig:autocorrelation_kappa}a shows that for serial execution and $\kappa = 10^{-4}$ the autocorrelation drops to zero after $\approx$30 MC routines. For the same value of $\kappa$, the parallel algorithm decorrelates samples more quickly. When $\kappa$ is increased to $10^{-3}$, correlations die out in fewer than 5 MC routines.

\textbf{ALT TEXT}
Figure S3a shows the change in average titanium concentration measured in VC-SGC MCMD of Hf--Ti at 1500\,K as the variance-constraint parameter $\kappa$ is varied between $10^{-4}$ and 1. At the same time, the number of MPI ranks takes values of 1 (serial algorithm), 2, 4, and 8. Parallel execution with 2, 4, and 8 ranks fixes the composition exactly to the target value of 0.5, while the serial algorithm requires a $\kappa$ of at least $10^{-3}$. Figure S3b shows the change in MC acceptance probabilities for the same simulations as $\kappa$ and the number of MPI ranks are changed. The serial algorithm reaches the highest acceptance probability of $\approx0.48$ around $\kappa = 10^{-3}$ before dropping with increasing $\kappa$. The probability curves for 2, 4, and 8 ranks have similar shapes. The highest probability of $\approx0.44$ occurs at $10^{-4}$, and it monotonically decreases with increasing $\kappa$.

\section{Computational details} \label{sect:computational_details}
All MCMD simulations in this work used a Nosé--Hoover thermostat and barostat with damping times of 100 and 1000 time steps, respectively. For a time step of 0.002\,ps, the damping times amounted to 0.2 and 2\,ps. The variance-constraint parameter $\kappa$ was $10^{-3}$ for Hf--Ti (\cref{main-sect:HfTi}) and $0.5 \cdot 10^{-3}$ for Cu--Zr (\cref{main-sect:CuZr}). 

In the equilibrium Frenkel--Ladd free energy calculations of \cref{main-sect:HfTi} \cite{frenkel_new_1984}, the equilibrium lattice parameters were measured from 60\,ps of NPT molecular dynamics. Thermodynamic integration was carried out in the NVT ensemble at the equilibrium volume, adopting a scaling parameter mesh size of 0.02 and a Langevin thermostat with damping time of 0.1\,ps. Each intermediate state was equilibrated for 10\,ps, and sampling occurred for 50\,ps. The spring constants of the reference Einstein crystals were calculated from the mean-squared displacement \cite{freitas_nonequilibrium_2016}, yielding values of 2.89 and 1.46\,eV/{\AA}$^2$ for HCP and BCC Hf, respectively.

\textbf{ALT TEXT}
Figures S4a through S4d show the composition autocorrelation functions for two values of variance-constraint parameter $\kappa$ ($10^{-4}$ and $10^{-3}$) and a different number of MPI ranks (1, 2, 4, and 8 for figures a, b, c, and d, respectively). The change in the autocorrelation function is expressed in terms of the number of MC routines. For $\kappa = 10^{-4}$, the serial algorithm decorrelates samples in around 30 MC routines, while parallel execution decorrelates samples more quickly. For $\kappa = 10^{-3}$, samples become uncorrelated after around 5 MC routines regardless of the number of ranks.

\bibliographystyle{unsrt}
\bibliography{bibliography}